\documentclass[]{interact}

\usepackage{epstopdf}% To incorporate .eps illustrations using PDFLaTeX, etc.
\usepackage[caption=false]{subfig}% Support for small, `sub' figures and tables

\usepackage[doublespacing]{setspace}% To produce a `double spaced' document if required
\usepackage[numbers,sort&compress,merge]{natbib}% Citation support using natbib.sty
\bibpunct[, ]{[}{]}{,}{n}{,}{,}% Citation support using natbib.sty
\renewcommand\bibfont{\fontsize{10}{12}\selectfont}% Bibliography support using natbib.sty

\theoremstyle{plain}% Theorem-like structures provided by amsthm.sty

\theoremstyle{definition}

\theoremstyle{remark}

\usepackage{graphicx}
\usepackage{amssymb}
\usepackage{amsmath}
\usepackage[utf8]{inputenc}
\usepackage{cancel}
\usepackage[usenames,dvipsnames]{color}
\usepackage{psfrag}
\usepackage{epsfig}
\usepackage{bbm}
\usepackage{bm}
\usepackage{dsfont}
\usepackage{soul}
\usepackage{colonequals}
\usepackage{float}
\usepackage{enumitem}

\usepackage[colorlinks, allcolors = blue] {hyperref}
\makeatletter
\renewcommand*{\eqref}[1]{%
	\hyperref[{#1}]{\textup{\tagform@{\ref*{#1}}}}%
}
\makeatother

\newcommand{\bs}[1]{\boldsymbol{#1}} %% to make bold vectors
\newcommand{\makeAverageSystemDensitySymbol}{\bar{\rho}} %% rhoBar symbol

\begin{document}
%\linenumbers %%% turn on line numbering

%\articletype{ARTICLE TEMPLATE}% Specify the article type or omit as appropriate

\title{Freezing of a soft-core fluid in a one-dimensional potential:\\ Appearance of a locked smectic phase}

\author{
	\name{Alexander Kraft and Sabine H.~L.~Klapp}
	\affil{
		Institut f\"ur Theoretische Physik,
		Hardenbergstr.~36,
		Technische Universit\"at Berlin,
		D-10623 Berlin,
		Germany}
}

\date{\today}

\maketitle

\begin{abstract}
	We investigate the phase behaviour of a two-dimensional colloidal model system of ultra-soft particles on a substrate which varies periodically along one spatial direction.
	Our calculations are based on mean-field density functional theory for a system of particles interacting via an ultra-soft potential, that is, the generalized exponential model with index four (\mbox{GEM-4}). 
	For suitable substrate periodicities (with commensurability parameter $p=2$), we find a succession of phase transitions from a modulated liquid to a locked smectic and then to a locked floating solid phase.
	The appearance of a locked smectic phase is consistent with earlier theoretical predictions and experiments for freezing of more repulsive systems on structured surfaces (with $p=2$).
	However, the present ultra-soft system does not display re-entrant melting.
	We here investigate the details of the density distributions of the different phases, thereby supplementing earlier work on GEM-4 systems with $p=1$ [Phys. Rev. E \textbf{101}, 012609 (2020)].
	Interestingly, the observed succession of phase transitions can be triggered through different paths along which physical control parameters are changed.
\end{abstract}


%%%%%%%%%%%%%%%%%%%%%%%%%%%%%%%%%%%%%%%%%%%%%%%%%%%%%%%%%%%%%%%%%%%%%%%%%%%%%%%%%%%%%%%%%%%%%%%
\section{Introduction \label{SEC:INTRO}}

It is well established that the presence of structured surfaces can have a profound impact on the phase behaviour and particularly, the freezing transition of atomic, molecular, and colloidal fluids.
Typical effects are shifts of the freezing transition with respect to the corresponding bulk transition \cite{Alba_Simionesco_2006}, and a significant impact on the fluid's structure close to the walls \cite{schoen2007nanoconfined, Seemann1848}.
Examples include water at the inner surfaces of silica nanopores \cite{B010086M,doi:10.1002/cphc.200800616} or at graphene sheets \cite{PhysRevB.95.195414}, 
atoms between the structured surfaces of a surface force apparatus \cite{doi:10.1063/1.466668}, but also wetting of crystalline phases of colloids close to patterned substrates \cite{Esztermann_2005} and active Janus particles at chemically decorated surfaces~\cite{doi:10.1063/1.5091760}. In some (yet not all) cases, structured surfaces actually assist the adjacent fluid in developing a solid-like structure, that is, freezing is supported (relative to the bulk system) rather than suppressed.

Here we are interested in a seemingly "old" example of the first scenario, that is, the
freezing of a two-dimensional (2D) system of colloidal particles on a one-dimensional (1D) periodic substrate. This phenomenon, commonly denoted as laser-induced freezing (LIF), was first  discovered experimentally by Chowdhury, Ackerson, and Clark~\cite{Chowdhury1985} in a 2D monolayer of charged particles subject to a 1D periodic laser field. At low light intensities, i.e., low potential barriers $V_0$, and not too high average densities, the suspension forms a modulated liquid (ML) phase, characterized by an oscillatory density profile perpendicular to the stripes but full translational symmetry along the stripes. This changes at large light intensities, i.e., large $V_0$, where a "locked floating solid" (LFS) emerges. Here, the colloids are positionally locked perpendicular to the minima, but unlocked along them (thereby allowing the solid to "float" in one direction). This discovery motivated a 
series of studies by theory~\cite{Chakrabarti1994, Das1998, Das1999a, Frey1999, Radzihovsky2001, Rasmussen2002, Chaudhuri2004, Nielaba2004, Chaudhuri2006, Luo2009}, computer simulations~\cite{Loudiyi1992b, Chakrabarti1995, Das1999a, Das1999b, Das2001, Strepp2001, Strepp2002, Strepp2003, Chaudhuri2004,Chaudhuri2005, Chaudhuri2006, Buerzle2007, Luo2009} and experiments~\cite{Loudiyi1992a, Wei1998, Bechinger2000, Bechinger2001, Baumgartl2004}. From the theoretical side it turned out that mean-field like approaches (which are characterized by incorrect treatment of fluctuations) fail to predict the complete phenomenology \cite{Frey1999, Radzihovsky2001}. 
A major step towards a theoretical understanding of the full LIF scenario was provided by the work of Frey, Nelson, and Radzihovsky (FNR)~\cite{Frey1999, Radzihovsky2001} who extended the concept of dislocation-mediated melting in 2D described by KTHNY theory~\cite{Kosterlitz1973, Halperin1978, Nelson1979, Young1979} towards the presence of 1D periodic substrates~\cite{Bechinger2007}. Depending on the so-called commensurability parameter $p$ that depends on the substrate periodicity, $L_s$, and  determines the population of potential minima by particles, different phases with partial symmetry-breaking may arise.
Extensive numerical (Monte-Carlo) simulation studies~\cite{Strepp2001, Strepp2002, Strepp2003, Buerzle2007} later confirmed their results.

Until recently, studies of LIF were restricted to systems of particles with strongly repulsive interactions, although it is nowadays possible not only to fabricate soft particle-particle interactions~\cite{Liz-Marzan1996, Hoffmann2010, Ramli2013, Hayes2014}, but also to investigate their interaction with a substrate~\cite{Zaidouny2013, Schoch_Langmuir2014, Schoch_SoftMatter2014}.
This motivated us in an earlier study based on classical density functional theory \cite{Kraft2020a} to investigate the phenomenon of LIF in a system of particles interacting via an ultra-soft potential characterized by a finite value at zero separation, thus allowing for overlap. We studied this system on cosine and Gaussian substrates, focusing on the case $p=1$ (where each potential minimum is equally populated). Despite a mean-field like treatment, we could establish the occurrence of LIF and provide full phase diagrams in the planes spanned by the average density $\bar{\rho}$ and the parameters controlling the fluid-substrate potential. We also showed that LIF can be understood as a density-driven transition, thereby complementing the more traditional view where the control parameter is $V_0$.

In the present paper, we extend the methodology of \cite{Kraft2020a} to systems with commensurability parameter $p=2$, for which, in the ordered phase, the particle distribution in direction perpendicular to the substrate minima has periodicity $2L_s$. At $p=2$, the theory of Frey, Nelson, and Radzihovsky~\cite{Frey1999, Radzihovsky2001} predicts that the (re-entrant) melting of the LFS (with $p=2$) upon increasing $V_0$ occurs through two phase transitions with successive unbinding of dislocation pairs. First, unbinding of dislocation pairs with Burgers vectors parallel to the minima leads to a "locked smectic" (LSm) phase, which is liquid-like along the minima but still breaks the discrete symmetry of the substrate by populating only every second minimum equally. This is followed by an unbinding of dislocation pairs with Burgers vectors perpendicular to the minima, which eventually leads to a phase transition from the LSm into the ML phase.
Within the latter, the discrete substrate symmetry is restored, that is, the density profile displays modulations with periodicity $L_s$.
The emergence of a LSm phase upon melting the LFS was observed experimentally in a 2D colloidal system of charged polystyrene spheres~\cite{Baumgartl2004}, and it was
found in Monte Carlo simulation of hard discs~\cite{Buerzle2007}, It also appeared in a theoretical study of vortex systems in superconductors~\cite{Hu2005}.

In the present study we demonstrate that a LSm phase appears in systems of ultra-soft colloids. In particular, according to our mean-field density functional study, the LSm appears as an intermediate phase in between the ML and the LFS upon increasing either the potential barrier, or the average density, or by decreasing the available space within one minimum by manipulating the fluid-substrate potential.
We show this by analysing two-dimensional density profiles obtained by minimization of the grand canonical functional. The LSm is then identified by homogeneity along the minima, and periodicity $2L_s$ perpendicular to them. Due to the mean-field character of our approach, the occurrence of a LSm is indeed not an obvious result.
Different from other studies, however, we do not see re-entrant melting~\cite{Wei1998}, which is consistent with our previous study of systems at $p=1$~\cite{Kraft2020a}, but which is in contrast to observations in charged polystyrene spheres~\cite{Baumgartl2004} and for hard discs~\cite{Buerzle2007}.

The rest of this manuscript is organized as follows:
In Section~\ref{SEC:Theory} we introduce our 2D model system of ultra-soft particles, the two types of 1D periodic substrates considered, and the density functional treatment on which our work is based.
Numerical results from minimization of the density functional are presented in Section~\ref{SEC_Numerical_Results}.
In Section~\ref{SEC:Conclusion_and_Outlook}, we summarize and give an outline for future investigation, including preliminary results for new phases emerging at larger substrate periodicities.

%%%%%%%%%%%%%%%%%%%
\section{Theoretical framework \label{SEC:Theory}}

Our present study is based on the same type of model (ultra-soft colloids) and same method of investigation (classical density functional theory in mean-field approximation) as our earlier study on systems with commensurability parameter $p=1$~\cite{Kraft2020a}.  Therefore, we summarize only briefly the main points and refer the reader  for details to Ref.~\cite{Kraft2020a}.

\subsection{Model system}
We consider a 2D colloidal system (on the $x$-$y$ plane of the coordinate system) exposed to two variants of 1D periodic substrate potentials. The most simple variant is a harmonic (cosine) substrate potential, 
\begin{align}
	V_{\text{ext}}(\bs{r} ) =  \frac{V_0}{2} \cos\left( \frac{2 \pi}{L_{s}} x \right),
	\label{Eq_cosine_substrate}
\end{align}
with periodicity $L_{s}$ and amplitude~$V_0$, and the position vector $\bs{r} = (x,y) \in \mathbb{R}^2$.
Furthermore, we consider the Gaussian substrate 
\begin{align}
	V_{\text{ext}}(\bs{r} ) = \sum_{m \in \mathbb{Z}}  V_0 \exp\left( - \left( \frac{ x-m {L_{s}} }{ R_{g}} \right)^2 \right),
	\label{Eq_Gaussian_substrate}
\end{align}
where $R_{g}$ is a measure of the range of the Gaussian, and $m$ runs over all integer numbers.
Our motivation to introduce the Gaussian substrate is its tunability: 
It allows to effectively reduce the available space around the potential minima through the range $R_g$ of the Gaussian maxima. A comparison of both substrates can be found in Ref.~\cite{Kraft2020a}.

As for the colloidal system, we consider (for reasons outlined below and, in more detail, in Ref.~\cite{Kraft2020a}) a 2D system of ultra-soft particles, with the interaction given by the generalized exponential model of index $n$ (GEM-$n$),
\begin{align}
	\label{Eq_Interaction_Potential}
	V( |\bs{r}_1 - \bs{r}_2| ) = \epsilon \, \exp\left( - \left(\frac{  |\bs{r}_1 - \bs{r}_2|  }{R}\right)^n   \right).
\end{align}
In Equation~\eqref{Eq_Interaction_Potential}, $\bs{r}_1$ and $\bs{r}_2$ are the particle positions, $\epsilon$~is the interaction strength, and $R$ represents the range of the interaction.
As in Ref.~\cite{Kraft2020a}, we fix $n=4$ in Equation~\eqref{Eq_Interaction_Potential} throughout this work, and denote all length scales in units of $R$, the range of the particle interaction. 
The particle interaction strength is set to $\beta \epsilon = 1$, where $\beta = 1/ k_B T$ (with $k_B$ being Boltzmann's constant and $T$ being the temperature).

An important parameter in the context of LIF is the commensurability parameter $p$~\cite{Bechinger2007}.
	The commensurability parameter $p$~\cite{Bechinger2007} is given as the ratio
	\begin{equation}
		p = \frac{a^{\prime}_{\vec{m}}}{L_s}  = \frac{ |\bs{K}| }{ | \bs{G}_{\vec{m}} | }
		\label{Eq_Def_commensurability_parameter_p}
	\end{equation}
	between the distance  $a^{\prime}_{\vec{m}}$ between the lattice planes with Miller indices $\vec{m} =(m_1,m_2)$ (with $m_i \in \mathbb{Z}$) of the arising solid phase, and the substrate periodicity $L_s$.
	The second member of Equation~\eqref{Eq_Def_commensurability_parameter_p} expresses this ratio in Fourier space using reciprocal lattice vectors. Specifically, $\bs{K}$ denotes the dominant wave vector of the substrate potential (with $|\bs{K}| = 2 \pi / L_s$) and $\bs{G}_{\vec{m}} = m_1 \bs{G}_1 + m_2 \bs{G}_2$ denotes the reciprocal lattice vectors (with $|\bs{G}_{\vec{m}}| = 2 \pi / a^{\prime}_{\vec{m}}$).	
	Commensurability requires that the wave vector $\bs{K}$ of the substrate is equal to one of the reciprocal lattice vectors $\bs{G}_{\vec{m}}$~\cite{Bechinger2007}.
	We here rather focus on the real space representation, where commensurability requires that one of the lattice planes of the arising solid coincides with the substrate minima. 
	The case $p = N$ (with $N$ being a natural number) then corresponds to a situation where each ($p=1$) or every $p$-th ($p>1$) minimum is equally populated.
In Ref.~\cite{Kraft2020a}, we have considered the case $p=1$, where the locked floating solid into which the (modulated) liquid freezes is characterized by lattice sites in \textit{each} minimum of the periodic substrate. 
This situations occurs most likely when the formed (locked floating) solid has the same lattice constant $a$ as the bulk solid, and the relative orientation with respect to the substrate is a primary orientation~\cite{Bechinger2007}.
For the present GEM-4 potential, the (2D) bulk lattice constant is $a/R \approx 1.4$ (see e.g. Ref.~\cite{Kraft2020a}), yielding $L_s/R = \sqrt{3}a/2R \approx 1.2$ as an optimal value for $p=1$.
Phases with commensurability parameter $p=2$, in particular the LSm ($p=2$) and the LFS ($p=2$), are expected at substrate periodicity $L_s = a \sqrt{3}/4$~\cite{Bechinger2007}. 
This value was also used in the studies of the LSm ($p=2$) for charged polystyrene spheres~\cite{Baumgartl2004} and for hard discs \cite{Buerzle2007}.
For the present GEM-4 system, we obtain with $a/R \approx 1.4$, a dimensionless value of $L_s /R \approx 0.6$ as an optimal choice for the case $p=2$.

\subsection{Density functional theory \label{SUBSEC:Model_and_density_functional_theory_subsection_DFT}}

To calculate the equilibrium density profile, $\rho_{\text{eq}}(\bs{r})$, we use classical DFT~\mbox{\cite{Evans1979,Evans1992}}.
The main idea is that $\rho_{\text{eq}}(\bs{r})$ minimizes the grand potential functional 
\begin{align}
	\Omega[\rho] &= F[\rho] +  \int d\bs{r} \rho(\bs{r}) V_{\text{ext}}(\bs{r})  - \mu \int d\bs{r} \rho(\bs{r}) 
	\label{Eq_grand_potential_functional}
\end{align}
with chemical potential~$\mu$, external potential~$V_{\text{ext}}(\bs{r})$, and the intrinsic Helmholtz free energy functional \mbox{$F[\rho] = F_{\text{id}}[\rho] + F_{\text{exc}}[\rho]$}.
The ideal gas contribution of $ F[\rho]$ is known exactly,
\begin{subequations}
	\begin{align}
		F_{\text{id}}[\rho] &= k_B T \int d\bs{r} \rho(\bs{r}) \left[ \ln(\Lambda^2 \rho(\bs{r}))-1 \right], 
		\label{Eq_ideal_gas_free_energy_functional} \\
		\intertext{where $\Lambda$ is the de Broglie wavelength.
			The excess free energy~$F_{\text{exc}}$ describes the impact of the interactions between particles, and has to be approximated for most types of interactions.
			Consistent with our earlier study~\cite{Kraft2020a}, we use the mean-field (MF) approximation for $F_{\text{exc}}$ that is well established for the description of ultra-soft particles at high density~\cite{Likos2001},}
		F_{\text{exc}}[\rho] &=  \frac{1}{2} \int d\bs{r} \int d\bs{r}' \big[ \rho(\bs{r})  V(\bs{r}-\bs{r}')  \rho(\bs{r}') \big].
		\label{Eq_excess_free_energy_functional}
	\end{align}
	\label{Eq_intrinsic_free_energy_functional}
\end{subequations}
The high accuracy of the mean-field approximation for different types of ultra-soft particles was frequently demonstrated
\cite{Lang2000, Louis2000PhysRevE,Archer2004,Likos2007, Archer2014,Mladek2006, Mladek2007,Likos2007} (see Ref.~\cite{Kraft2020a} for a more detailed description). 
Apart from the direct connection to particle interactions, a further major benefit of the DFT approach from a practical point of view is the possibility of an unconstrained (numerical) minimization, in which no \textit{a priori} information of the spatial form of $\rho_{\text{eq}}(\bs{r})$ is assumed. 
All of the results in the present paper are based on such an unconstrained minimization.
We note that using a constrained minimization (using, e.g., arrays of Gaussian peaks to describe the density profile in solid-like phases) one could potentially not only miss details of the phases, but even entirely miss phases which are not covered by an \textit{a priori} prescribed ansatz.
While this is true in general, it seems particularly relevant for the system at hand.
Indeed, in the outlook (see Section~\ref{SEC:Conclusion_and_Outlook}), we show an example of a phase that we probably would have missed when using a prescribed ansatz rather than unconstrained minimization.

The minimization of Equation~\eqref{Eq_grand_potential_functional} leads to the Euler-Lagrange equation,
\begin{align}
	\rho_{\text{eq}}	(\bs{r})=\Lambda^{-2} 
	\exp\left[\beta \mu  -\beta V_{\text{ext}}(\bs{r}) - \beta \left.\frac{\delta F_{\text{exc}}[\rho]}{\delta \rho(\bs{r})}\right|_{\rho_{\text{eq}}} \right].
	\label{Eq_Euler_Lagrange}
\end{align}
Similar to our previous work~\cite{Kraft2020a}, we solve Equation~\eqref{Eq_Euler_Lagrange} self-consistently using (numerical) fixed-point iteration~\cite{Hughes2014} at given temperature, interaction parameters, and given average density~$\bar{\rho} = \langle N \rangle / (L_x L_y)$ (where $\langle N \rangle$ is the average particle number related to the chemical potential~$\mu$), and with periodic boundary conditions in both directions. 
	The numerical minimization closely follows the general scheme as e.g. presented in Ref.~\cite{Hughes2014}: The density profile $\rho(\bs{r})$ is discretized on a set of grid points with discretization $dx$ and $dy$, which yields a discretized density profile $\rho_{\bs{i}}$ with grid indices $\bs{i} = (i_x, i_y)$. 
	The (discretized) Equation \eqref{Eq_Euler_Lagrange} is then used as a fixed-point equation $\rho_{\bs{i}}^{n+1} = f[\rho_{\bs{i}}^{n}]$, where $f$ denotes the right hand side of Equation~\eqref{Eq_Euler_Lagrange}, $n$ is an iteration index of the fixed-point iteration. and iteration steps $n$ are done until the density profile $\rho_{\bs{i}}$ converges.
	For numerical stability reasons~\cite{Hughes2014}, the previous density profile $\rho_{\bs{i}}^{n}$ is mixed with the new density profile (as obtained from $f[\rho_{\bs{i}}^{n}]$) with a mixing parameter~$\alpha$ (typically $\alpha \in [0.01:0.1]$) to  obtain the next iteration step $\rho_{\bs{i}}^{n+1} = \alpha f[\rho_{\bs{i}}^{n}] + (1-\alpha) \rho_{\bs{i}}^{n}$.
	Other technical details	such as choice of discretization and initial condition are described in Appendix A of Ref.~\cite{Kraft2020a}.
From the density profile $\rho_\text{eq}(\bs{r})$ that we numerically obtain at given $\bar{\rho}$, the associated chemical potential $\mu$ is given through integration of Equation~\eqref{Eq_Euler_Lagrange} as 
\begin{align}
		\beta \mu = \ln\left(\Lambda^2 \bar{\rho} \right) - \ln\left( \frac{\int d\bs{r} \exp\left[-\beta V_{\text{ext}}(\bs{r}) - \beta \left.\frac{\delta F_{\text{exc}}[\rho]}{\delta \rho(\bs{r})}\right|_{\rho_{\text{eq}}} \right]}{L_x L_y}\right).
		\label{Eq_beta_mu_from_density_profile}
\end{align}

For the current study, we made a simple extension to speed up the numerical minimization.
We used two mixing parameters $\alpha_1, \alpha_2> \alpha_1$ instead of the mixing of density profiles with one mixing parameter $\alpha$ (being constant and fixed throughout the numerical calculation, see e.g. Ref.~\cite{Hughes2014} for an introduction). Specifically, after 1000 iterations with mixing parameter~$\alpha_1$, the mixing alternates between the two values of the  mixing parameters $\alpha_i$ where $\alpha_1$ remains fixed and $\alpha_2$ is adjusted automatically. The initially provided value of $\alpha_2$ is used as an maximum value for the mixing parameter $\alpha_2$ and can never be exceeded. 
The mixing parameter $\alpha_2$ is increased by a factor of 1.05, if the value of the grand potential $\Omega[\rho]$ [see Equation~\eqref{Eq_grand_potential_functional}] decreased over the last 500 iterations, and $\alpha_2$ is decreased by a factor of $0.8$ otherwise. We found that this simple extension greatly accelerated the numerical calculations, while the results were unchanged compared to those with simple mixing.

%%%%%%%%%%%%%%%%%%%%%%%%%%%%%%%%%%%%%%%%%%%%%%%%%%%%%%%%%%%%%%%%
\section{Numerical Results\label{SEC_Numerical_Results}}

In this section we present our numerical results obtained by minimization of the grand potential $\Omega$ [see Equation~\eqref{Eq_grand_potential_functional}] at various average densities~$\makeAverageSystemDensitySymbol$ and various parameters of the external potential $V_0$ or $R_g$ (see Equations~\eqref{Eq_cosine_substrate} and \eqref{Eq_Gaussian_substrate} for the cosine and Gaussian substrate, respectively).
We observe three types of phases; the ML, LSm, and LFS.
In Section~\ref{Sec:Characteristics_of_the_ML_LSm_LFS_phases}, we first discuss characteristic features of each phase as reflected by the density distribution.
In Section~\ref{Sec:Phase_Diagrams}, we then present full phase diagrams involving different control parameters.

\subsection{Characteristics of the different phases\label{Sec:Characteristics_of_the_ML_LSm_LFS_phases}}
In earlier studies of LIF, the different phases have often been identified by pair correlation functions (see e.g. Refs.~\cite{Buerzle2007, Baumgartl2004, Bechinger2007}), or the Fourier transformed density~\cite{Buerzle2007}. In the present mean-field DFT study, we rather investigate directly the density distributions in real space, which are a direct results of our calculations.
We start by discussing the density profiles on the cosine substrate~[see Equation~\eqref{Eq_cosine_substrate}] at fixed average system density $\makeAverageSystemDensitySymbol R^2= 4.5$ (i.e., far below the bulk freezing threshold) and various potential amplitudes $V_0$.
Given that the present substrate varies along the $x$-direction, we categorize the density profiles according to two criteria:
(i)~Whether they show the discrete symmetry of the substrate with periodicity $L_s$ along the $x$-direction, or rather twice $L_s$, and~(ii) whether they are homogeneous or inhomogeneous along the $y$-direction.

\begin{figure*}
	\centering	
	\includegraphics[width=0.95\textwidth]{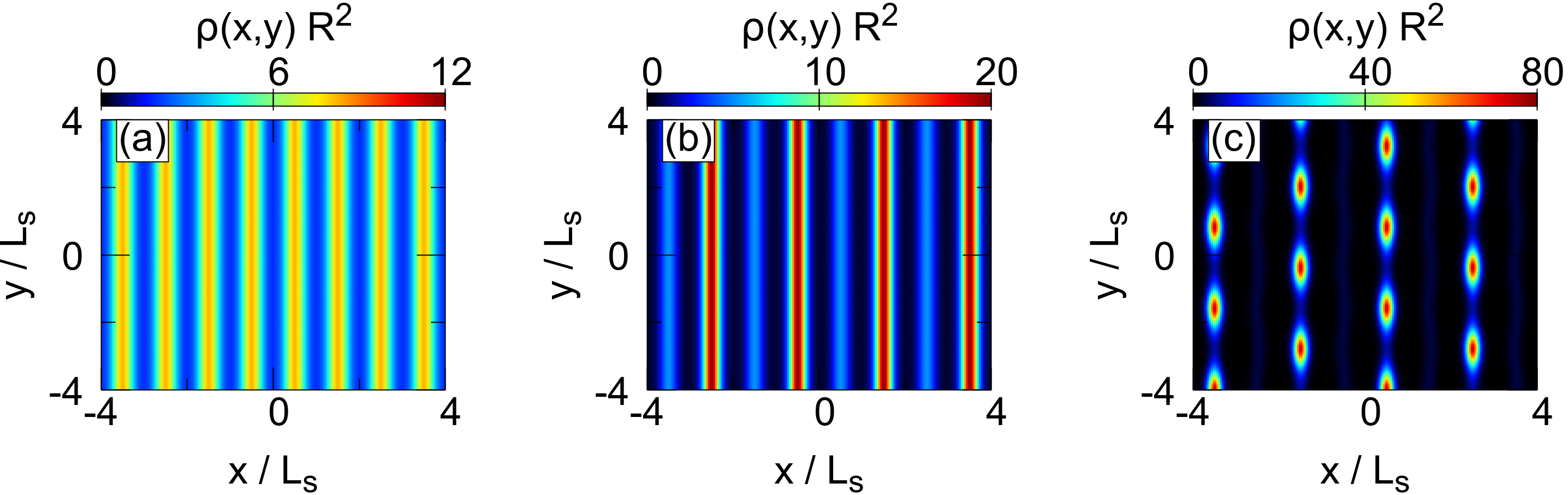}

	\caption{Representative density profiles $\rho(x,y)$  for different values of the potential amplitude $V_0$ on the cosine substrate [see Equation~\eqref{Eq_cosine_substrate}]. 
		Figure parts show (a)~the modulated liquid phase ($\beta V_0 = 1.5$), (b) the locked smectic phase ($\beta V_0 = 3$), and (c)~the locked floating solid phase ($\beta V_0 = 5$).
		In all parts, the average density is $\bar{\rho} \, R^2= 4.5$, and the substrate periodicity is $L_s / R = 0.6$.
	}
	\label{Fig_exemplary_density_profiles_rho=4.5}
\end{figure*}

Representative density profiles are shown in Figure~\ref{Fig_exemplary_density_profiles_rho=4.5}:
At low values of $V_0$, the modulated liquid phase (ML) arises [see Figure~\ref{Fig_exemplary_density_profiles_rho=4.5}(a)], where $\rho(\bs{r})$  varies only along~$x$ and displays the substrate periodicity~$L_s$. 
At intermediate $V_0$, the obtained density profiles reflect a symmetry-breaking of the discrete substrate symmetry [see~Figure~\ref{Fig_exemplary_density_profiles_rho=4.5}(b)].
They exhibit a periodicity $2 L_s$ along the $x$-direction, but are still constant along the $y$-direction. 
Given these features, we identify this state as a LSm phase, which has been previously observed in other colloidal systems, such as charged polystyrene spheres~\cite{Baumgartl2004} and hard discs~\cite{Buerzle2007}.
Moreover,  we find that the LSm phase that we observe in our calculations is characterized by different densities (i.e., different population of particles) in adjacent minima.
Upon further increase of $V_0$, the obtained density profiles are not only symmetry-broken (with respect to the substrate) in $x$-direction, but also are inhomogeneous in $y$-direction. Specifically, one observes hexagonal order.
Due to this, we identify this phase as the locked floating solid (LFS) phase~\cite{Bechinger2007} at $p=2$ [see~Figure~\ref{Fig_exemplary_density_profiles_rho=4.5}(c)]. 
Furthermore, we observe that the lattice sites of the LFS are located in the highly populated minima of the former LSm phase, and that the low density regions are nearly depleted of particles.
Taken together, we observe (at fixed $\bar{\rho} R^2$) the sequence ML-LSm-LFS upon increasing $V_0$ on a substrate with $p=2$. We note that the very appearance of the LSm is consistent with earlier studies~\cite{Baumgartl2004,Buerzle2007}, the order of the sequence of the transitions upon increasing $V_0$ is somewhat different.

\begin{figure}
	\centering
	
	\includegraphics[width=0.8\textwidth]{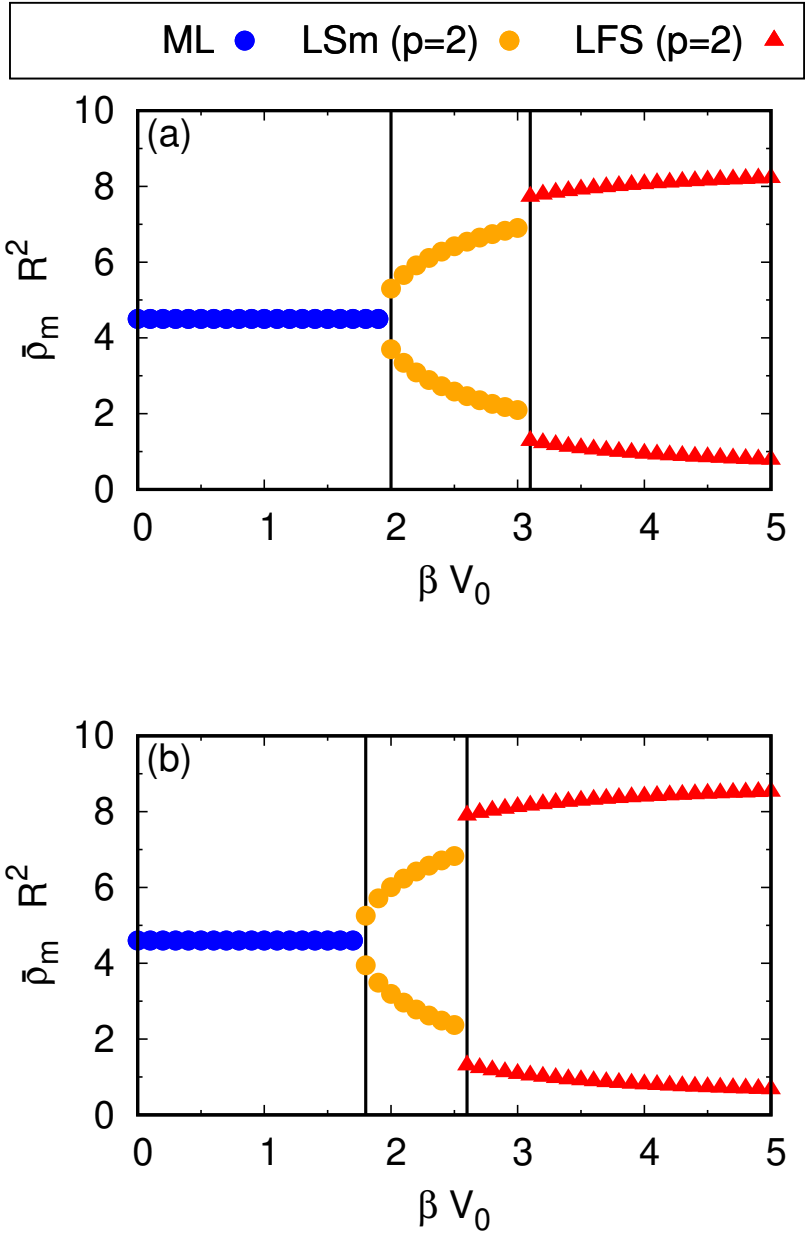}		
	
	\caption{Average density $\bar{\rho}_{m}$ in adjacent minima (say $x_{\text{min}}$ and $x_{\text{min}} + L_s$) [see Equation~\eqref{Eq_def_rho_modulation}] as function of $V_0$ on the cosine substrate ($p=2$). 
		In the modulated liquid phase (ML) both values of $\bar{\rho}_{m}$ in adjacent minima are identical (at given $V_0$).
		For the locked smectic phase (LSm) and the locked floating solid phase (LFS), two distinct values of $\bar{\rho}_{m}$ in the adjacent minima are obtained (at given $V_0$) corresponding to the alternating high and low density regions of the LSm and LFS in Figures~\ref{Fig_exemplary_density_profiles_rho=4.5}(b) and (c).
		The substrate periodicity is $L_s / R = 0.6$ for both figure parts.
		In part (a) the average density is $\bar{\rho} \, R^2= 4.5$ as in Figure~\ref{Fig_exemplary_density_profiles_rho=4.5}. 
		The left black solid line corresponds to the numerically obtained phase transition from the ML to the LSm ($p=2$) at $\beta V_0 = 2.0$.
		The right black solid line corresponds to the numerically obtained phase transition from the LSm ($p=2$) to the LFS ($p=2$) at $\beta V_0 = 3.1$.
		Part (b) shows the same analysis as in (a) but for a slightly higher average density $\bar{\rho} \, R^2= 4.6$.		
		The black solid lines have the same meaning as in (a) and are at $\beta V_0 = 1.8$ and $\beta V_0 = 2.6$.
	}

	\label{Fig_rhoBar_m_vs_V_0}
\end{figure}

\begin{figure}
	\centering
	
	\includegraphics[width=0.7\textwidth]{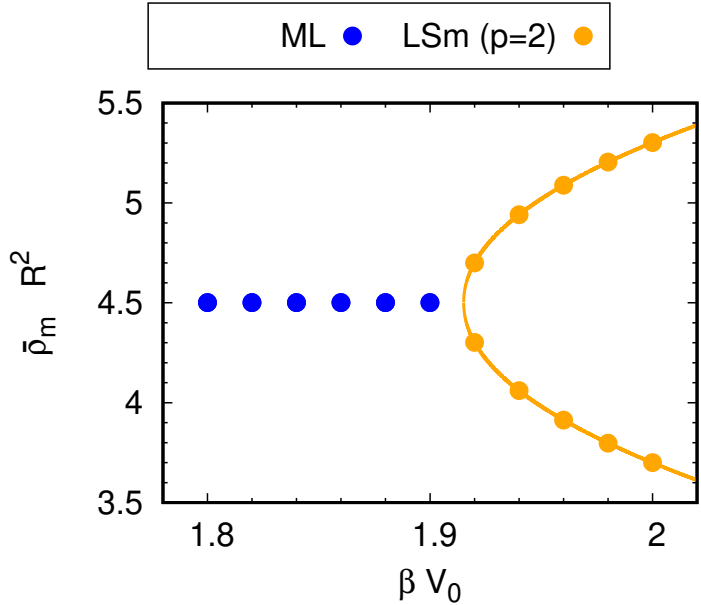}		
	
	\caption{
	Same as Figure \ref{Fig_rhoBar_m_vs_V_0}(a), but close to the modulated liquid (ML) - locked smectic (LSm) phase transition. The solid curve is a fit to the data points for the locked smectic phase as described in the main text.
	}

	\label{Fig_rhoBar_m_vs_V_0_Zoom_in_on_ML_LSm_transition}
\end{figure}

After demonstrating the existence of the intermediate LSm phase (as exemplarily presented in Figure~\ref{Fig_exemplary_density_profiles_rho=4.5}), we proceed by analysing our DFT data quantitatively.
In particular, to further investigate the symmetry breaking of the discrete substrate symmetry, we consider the average density $\bar{\rho}_{m}$ in one modulation of the substrate,
\begin{align}
	\bar{\rho}_{m} = \frac{1}{L_y L_s} \int\limits_{-L_y/2}^{L_y/2} dy  \int\limits_{x_{\text{min}} -  \frac{L_{s}}{2} }^{x_{\text{min}} +  \frac{L_{s}}{2} } dx\, \rho(x,y),
	\label{Eq_def_rho_modulation}
\end{align}	
and compare values of $\bar{\rho}_{m}$ in adjacent minima (say $x_{\text{min}}$ and $x_{\text{min}} + L_s$).
Thus, in addition to direct visual inspection of the density profile (see Figure~\ref{Fig_exemplary_density_profiles_rho=4.5}), we identify a broken translational symmetry between neighbouring minima based on the average density $\bar{\rho}_{m}$ in one modulation of the substrate.
When the discrete symmetry is broken, this leads to two distinct values of $\bar{\rho}_{m}$ in adjacent minima $x_{\text{min}}$ for the LSm and the LFS.
Figure~\ref{Fig_rhoBar_m_vs_V_0} shows values of $\bar{\rho}_{m}$ for various values of the potential amplitude~$V_0$ (at two values of the average density).
We clearly observe that a transition from a ML (characterized by one value of $\bar{\rho}_m$) to a LSm phase is accompanied by a splitting of $\bar{\rho}_m$ into two values related to adjacent minima.
The splitting arises without notable jumps, which indicates a continuous phase transition.
A more detailed view of the transition region ML-LSm is given in Figure~\ref{Fig_rhoBar_m_vs_V_0_Zoom_in_on_ML_LSm_transition}. 
We found that the data points for the LSm are well represented by a fitting function $\bar{\rho} + C (\beta V_0 -  \beta V_{0,c})^\nu$, where $C$ is a proportionality constant and $\nu$ denotes the critical exponent.
In particular, we obtained $\nu \approx 0.49$,	$\beta V_{0,c} \approx 1.915$, and $C  R^2\approx 2.7$ (with fitting errors below 1 \% for all quantities and numerical values slightly depending on the exact details of the fit).	
We thus find that the critical exponent $\nu$ is close to the value of 1/2, as is typical for a mean-field system~\cite{HansenMcDonald, Reichl1998modern}.	%
In contrast, upon increasing $V_0$, we see that the transition from the LSm to the LFS is accompanied by a discontinuous jump in the $\bar{\rho}_m$ values.
We note in passing that the low density region in the LFS is not completely depleted of particles, that is, $\bar{\rho}_m$ is clearly non-zero. Thus one may imagine that the low density regions mediate the interaction between the neighbouring high density regions in which the lattice sites are located.

Furthermore, it is interesting to consider the width of the region where the LSm appears as an intermediate phase between the ML and the LFS. In particular, we are interested in the sensitivity with respect to the density $\bar{\rho}$. We have thus repeated the same analysis for a slightly higher average density $\bar{\rho} \, R^2= 4.6$ than the previously chosen value $\bar{\rho} \, R^2= 4.5$. Results for $\bar{\rho}_m$ are shown in Figure~\ref{Fig_rhoBar_m_vs_V_0}(b). Despite the minor change in density~$\makeAverageSystemDensitySymbol$ (an increase by around 2.2\%), the range of $V_0$ values in which the LSm arises is drastically reduced (a decrease by roughly 27 \%).  In summary, we find that the closer the density $\makeAverageSystemDensitySymbol$ is to the bulk freezing density $\makeAverageSystemDensitySymbol_f$ (with $\bar{\rho}_f R^2 = 5.48$ according to \cite{Archer2014}), the smaller is the width of the intermediate LSm.
We will also see this explicitly in the phase diagrams presented in Section~\ref{Sec:Phase_Diagrams}.

To close this section, we turn our attention to the transition region between the LSm and LFS which (as seen in Figure~\ref{Fig_rhoBar_m_vs_V_0}) is characterized by a discontinuous behaviour of $\bar{\rho}_m$.
We repeatedly calculated density profiles in the vicinity of the transition (for slightly perturbed initial conditions) and observed a bistability, in the sense that calculated density profiles were either related to the LSm or the LFS phase.
The numerical DFT calculations thus hint at a coexistence between the LSm and LFS, and therefore a first order phase transition between these two phases.
We note that the data in the transition region in Figure~\ref{Fig_rhoBar_m_vs_V_0} was obtained by manually selecting the profile with minimal grand potential $\Omega$.
This procedure, however, becomes quite unfeasible when scanning entire phase diagrams.
We will return to this question of the order of the phase transitions in Section~\ref{Sec:Phase_Diagrams}.

%%%%%%%%%%%%%%%%%%%%%%%%%%
\subsection{Phase diagrams\label{Sec:Phase_Diagrams}}

\begin{figure*}
	\centering
	\includegraphics[width=0.95\textwidth]{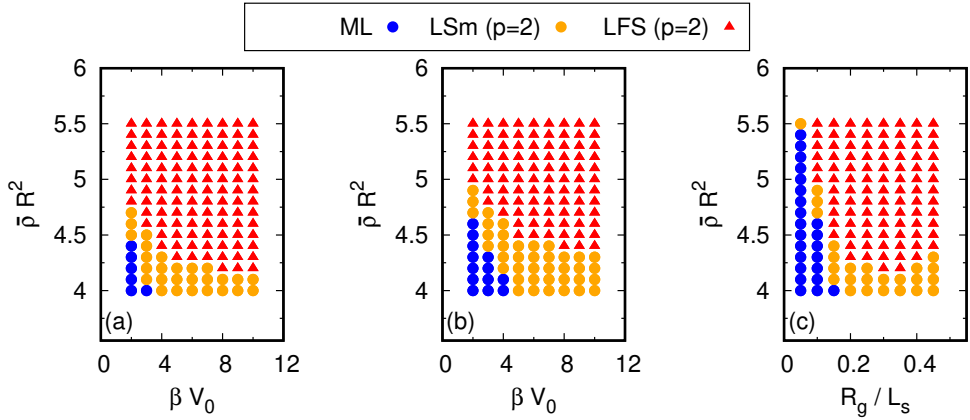}
	
	\caption{Phase diagrams obtained from DFT minimization for various average densities~$\bar{\rho}$ on 
		(a) the cosine substrate for varying potential amplitude~$V_0$, 
		(b) the Gaussian substrate for varying $V_0$ and fixed range $R_{g}$ ($R_{g}/L_s = 0.2$),  and 
		(c) the Gaussian substrate for varying $R_g$ and fixed $V_0$ ($\beta V_0 = 10$).
		The symbol type encodes whether the obtained phase is a modulated liquid (ML), a locked smectic (LSm) or a locked floating solid (LFS). 
		The substrate periodicity is $L_s / R= 0.6$.
		In the parameter ranges $\beta V_0 < 2$ and $R_g/L_s < 0.05$ [i.e. in the ranges left of the data points in (a)-(c)], we have performed test calculations suggesting that the stable phase is a ML. We also note that the bulk coexistence densities for the liquid and solid phase are $\bar{\rho}_l R^2 = 5.48$ and $\bar{\rho}_s R^2 = 5.73$~\cite{Archer2014}.
		}	
	
	\label{Fig_CompareParameterScanWithPrediction_L_s=0.60}	
\end{figure*}

In this section, we present an overview of the phase behaviour of the GEM-4 system on the cosine and the Gaussian substrate [see Equations~\eqref{Eq_cosine_substrate} and \eqref{Eq_Gaussian_substrate}] with $p=2$.
To this end, we have scanned large portions of the phase diagram on both substrates.
We indeed found the same sequence of ML-LSm-LFS phase transitions for different physical scenarios, as the phase diagrams in Figure~\ref{Fig_CompareParameterScanWithPrediction_L_s=0.60} reveal:
(i)~At constant substrate amplitude $V_0$, upon increasing the average density $\bar{\rho}$ of the system.
(ii)~At constant average density $\bar{\rho}$, upon increasing the substrate amplitude $V_0$.
(iii)~At constant $\bar{\rho}$ and constant $V_0$, upon reducing the available space for particles (that is, by increasing the range of the Gaussian maxima $R_g$ on a Gaussian substrate).
We observe this same sequence of ML-LSm-LFS phase transitions for two different types of substrates (cosine and Gaussian), thus demonstrating that it is not a peculiarity of the substrate. 
A common feature of all three diagrams in Figure~\ref{Fig_CompareParameterScanWithPrediction_L_s=0.60} is that the range of control parameters (i.e., $V_0$ or $R_g$) where the LSm arises becomes narrower with increasing average density (as already indicated at the end of Section \ref{Sec:Characteristics_of_the_ML_LSm_LFS_phases}).

We now focus in more detail on the diagrams obtained upon variation of $V_0$ [see Figures ~\ref{Fig_CompareParameterScanWithPrediction_L_s=0.60}(a) and (b)].
Here, the LSm phase appears either in between the ML and LFS phase (high densities) or as the (only) stable phase at large $V_0$ in the range considered by us (low densities).
It is indeed unclear whether the LSm phase in the present system will eventually freeze into a LFS at sufficiently large values of $V_0$. In fact, a similar observation has been made in Monte Carlo simulations of hard discs~\cite{Buerzle2007}. There, the phase diagram shows that for values of the potential amplitude $V_0$ as large as $\beta V_0 = 10000$, there is a range of densities for which the LSm phase remains the stable phase and does not freeze into a LFS.
Furthermore, the possibility of a LSm remaining the stable phase at large $V_0$ is also in agreement with the theoretical prediction of Frey, Nelson, and Radzihovsky~\cite{Frey1999, Radzihovsky2001} (see, in particular, Figure 3(a) in \cite{Frey1999}).

Furthermore, interestingly, we do not observe a re-entrant  melting, i.e., a  transition from the more ordered 
LFS phase to the less ordered LSm or ML phase upon increase of $V_0$.
In this regard, our results differ from experimental results for charged polystyrene spheres~\cite{Baumgartl2004} and MC data for hard discs \cite{Buerzle2007} at $p=2$, where re-entrant melting occurs and causes an "up-bending" of the transition curves at large $V_0$ (see e.g. Ref.~\cite{Buerzle2007}).
We have seen a similar discrepancy in our previous mean-field-DFT study of the case $p=1$ \cite{Kraft2020a}, where we didn't find re-entrant melting of the LFS ($p=1$) solid upon increase of $V_0$, contrary to corresponding findings of more repulsive systems at $p=1$ in the literature.
Whether or not the absence of re-entrant melting is a feature of the ultra-soft system considered here, or an artefact of our mean-field treatment, remains to be explored.
Turning now to the phase diagram obtained through variation of $R_g$ [see Figure ~\ref{Fig_CompareParameterScanWithPrediction_L_s=0.60}(c)], we indeed find a re-entrant melting
(very similar to the same system at $p=1$ \cite{Kraft2020a}).
However, here the physical reason is different: upon increasing $R_g$ to large values, the overlap of neighbouring Gaussian maxima of the substrate becomes more and more significant. This causes a reduction of the effective barrier felt at a potential minimum, which eventually leads to the up-bending visible in Figure~\ref{Fig_CompareParameterScanWithPrediction_L_s=0.60}(c). 

\begin{figure*}
	\centering
	\includegraphics[width=0.95\textwidth]{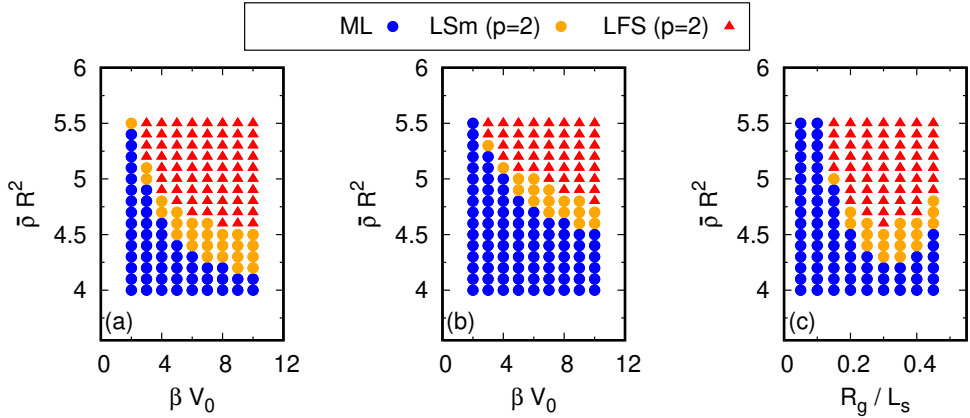}
	
	\caption{Same as Figure~\ref{Fig_CompareParameterScanWithPrediction_L_s=0.60}, but for substrate periodicity $L_s / R= 0.55$.
	}

	\label{Fig_CompareParameterScanWithPrediction_L_s=0.55}	
\end{figure*}

So far we have focused on one value of the substrate periodicity. 
It is also interesting to study the influence on the substrate periodicity being slightly away from the value of $L_s/R = 0.6$ where the LSm and the LFS fit perfectly on the substrate and are thus, to some extent, expected~\cite{Bechinger2007}.
To this end, we show in Figure~\ref{Fig_CompareParameterScanWithPrediction_L_s=0.55} results for a slightly smaller value $L_s/R = 0.55$.
We find the same phenomenology as described in cases (i)-(iii) (see first paragraph of Section~\ref{Sec:Phase_Diagrams}), but with all transitions being shifted to higher values of the average density $\bar{\rho}$.
We also performed some calculations for $L_s/R=  0.65$, but did not observe a LSm ($p=2$) phase between the ML and the LFS ($p=2$). These exploratory calculations rather indicated that the ML first transforms in a LFS with each minimum being equally populated ($p=1$), followed by a transition into a LFS ($p=2$) phase.
However, we did not pursue this further, as it was not the focus of our current work.

\begin{figure*}
	\centering
	\includegraphics[width=0.95\textwidth]{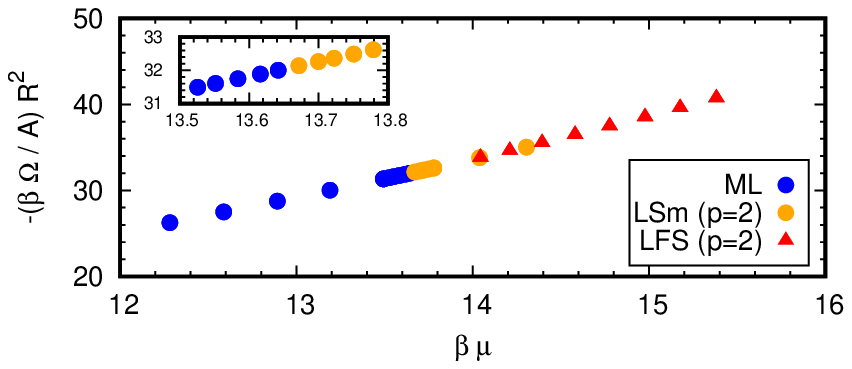}

	\caption{
		Grand potential $\Omega$ versus the chemical potential $\mu$ on the cosine substrate at $\beta V_0 = 2$. The symbol encoding for the phases is the same as in Figure~\ref{Fig_CompareParameterScanWithPrediction_L_s=0.60}(a).
		The substrate periodicity is $L_s / R= 0.60$.
		In the inset, we provide a closer view of the ML-LSm transition.
	}
	
	\label{Fig_investigate_order_of_phase_transitions_cosine}	
\end{figure*}

Finally, we return to the question of the order of the ML-LSm-LFS phase transitions (within the present mean-field DFT approach) at the optimal substrate periodicity $L_s/R = 0.6$.
In Figure~\ref{Fig_investigate_order_of_phase_transitions_cosine}, we plot the grand potential $\Omega$ versus chemical potential $\mu$ on the cosine substrate. For this we make a cut along Figure~\ref{Fig_CompareParameterScanWithPrediction_L_s=0.60}(a) at fixed value of the potential amplitude ($\beta V_0 = 2$) and vary the density $\bar{\rho}$.
The grand potential~$\Omega$ is obtained through minimization of Equation~\eqref{Eq_grand_potential_functional} and the chemical potential $\mu$ follows as the associated Lagrange parameter for given system density~$\bar{\rho}$ through Equation~\eqref{Eq_beta_mu_from_density_profile}.
The slope of the curve $\Omega(\mu)$, which corresponds to the overall density, appears to be constant at the ML-LSm transition (see inset of Figure~\ref{Fig_investigate_order_of_phase_transitions_cosine}) suggesting that this transition is continuous with respect to $\bar{\rho}$.
At this point it is also worth to recall the results in Figures~\ref{Fig_rhoBar_m_vs_V_0} and \ref{Fig_rhoBar_m_vs_V_0_Zoom_in_on_ML_LSm_transition}, where we found that the order parameter $\bar{\rho}_m$ as well changes continuously at the ML-LSm transition.
Regarding the LSm-LFS transition, the results in Figure~\ref{Fig_investigate_order_of_phase_transitions_cosine} (and additional calculations not shown here) indicate a slight change of slope of the curve, and furthermore there is some overlap of the two branches related to the LSm and the LFS phases.
This suggests that there are metastable regions, supporting the picture of a first order LSm-LFS phase transition.
In summary, our results from mean-field DFT point to a continuous ML-LSm phase transition and a first order LSm-LFS phase transition.

\section{Conclusion and Outlook\label{SEC:Conclusion_and_Outlook}}

In this work, we studied the phase behaviour of a colloidal model system of ultra-soft particles subjected to two variants of one-dimensional periodic substrates.
We here focused on systems characterized by a commensurability parameter $p=2$, thereby supplementing our previous analysis for $p=1$~\cite{Kraft2020a}.
Our results are based on classical density functional theory in the mean-field approximation, and we obtained the density profiles $\rho(\bs{r})$ by (unconstrained) minimization of the grand potential $\Omega$.

Most importantly, we found an intermediate locked smectic phase ($p=2$) between a modulated liquid and a locked solid phase ($p=2$). Such a phase was predicted theoretically based on an elastic Hamiltonian~\cite{Frey1999, Radzihovsky2001}, but has been observed, so far, only in experiments~\cite{Baumgartl2004} and MC simulations~\cite{Buerzle2007} of more repulsive systems.
A closer investigation of the locked smectic phase revealed that the breaking of the substrate periodicity is accompanied by a splitting of the density distribution into alternating high and low-density regions, thus creating a periodicity of $2 L_s$.
At sufficiently high potential amplitudes $V_0$,  the system freezes, and the former high density regions (of the locked smectic phase) contain lattice sites with hexagonal order; indicating a locked floating solid phase with $p=2$.

Performing extensive calculations for both, cosine and Gaussian substrates, we demonstrated that the appearance of the locked smectic phase is not a peculiarity of a specific shape of the substrate potential.
Rather, the observed sequence of transitions is robust in the sense that it appears through variation of different control parameters:
(i)~upon increase of $\bar{\rho}$ at constant $V_0$,
(ii)~upon increase of $V_0$ at constant $\bar{\rho}$,
and (iii)~upon reducing the available space for particles in the case of the Gaussian substrate.
Interestingly, we did not observe re-entrant melting~\cite{Wei1998}, that is, a transition from an ordered to a less ordered phase upon increase of $V_0$.
This is different from previous results for hard discs~\cite{Buerzle2007} and charged particles~\cite{Baumgartl2004}, but consistent with our earlier results for $p=1$~\cite{Kraft2020a}. 
Regarding the order of transitions, our mean-field DFT results indicate a continuous transition between the modulated liquid and the locked smectic, and a first order transition between the locked smectic and the locked floating solid.

We also studied, for a few cases, the influence of the substrate periodicity.
For a value slightly smaller than the optimal one, we found the same phenomenology, but a shift towards higher values of the average density $\bar{\rho}$.
More dramatic changes (with disappearance of the locked smectic phase) appear at a slightly larger value of $L_s/R$ compared to the optimal one. Here we have only briefly touched this issue, which would be an interesting aspect for future investigations.
Further, one could explore whether the approach suggested by us in Ref.~\cite{Kraft2020b}, which relies solely on bulk quantities, could be extended towards prediction of the phase boundaries at $p=2$.

\begin{figure*}
	\centering	
	\includegraphics[width=0.95\textwidth]{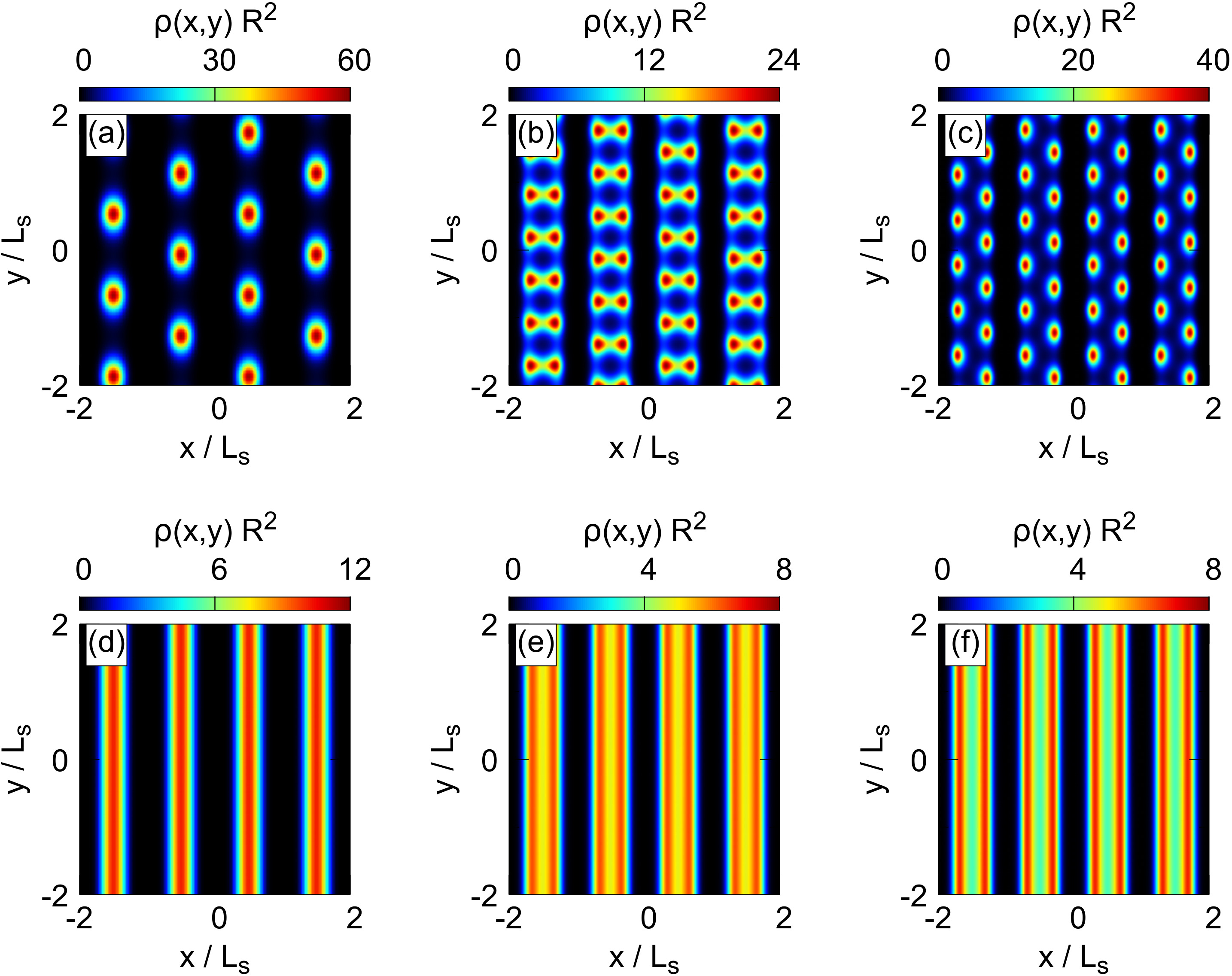}	
	
	\caption{Exemplary density profiles $\rho(x,y)$ obtained at large values of the substrate periodicity $L_s$ on the Gaussian substrate (at fixed $\beta V_0 = 10$, $R_g / L_s = 0.2$) [see Equation~\eqref{Eq_Gaussian_substrate}].
		Figure parts at the top (bottom) row are obtained at average system density $\bar{\rho} \, R^2= 5$ ($\bar{\rho} \, R^2=3$).
		Considering different substrate periodicities $L_s$, the following phases are obtained:
		(a) locked floating solid ($p=1$) ($L_s/R= 1.2$),
		(b) dumbbell solid ($L_s/R= 2.0$),
		(c) Two hexagonal lattice planes per minimum ($L_s/R= 2.2$),
		(d) modulated liquid with one maximum at the centre  ($L_s/R= 1.2$),
		(e) modulated liquid with two off-centre maxima ($L_s/R= 2.0$),
		(f) as (e) but now with clearly pronounced off-centre maxima in the density profile ($L_s/R= 2.2$).
	}
	\label{Fig_outlook_on_dumbbell_solid}
\end{figure*}

Yet another potentially interesting extension of the present work would be a numerical DFT study of much larger values of $L_s/R$.
Indeed, preliminary calculations for Gaussian substrates with $L_s/R$ in the range $1.2-2.2$ revealed a variety of rather exotic phases with density profiles shown in Figure~\ref{Fig_outlook_on_dumbbell_solid}. An intriguing example is the "dumbbell solid" in Figure~\ref{Fig_outlook_on_dumbbell_solid}(b), which is characterized by a double-peaked density distribution around each lattice site.
Other new solid phases (for different substrate periodicities $L_s$) at high density $\bar{\rho} R^2 = 5$ are shown in Figures~\ref{Fig_outlook_on_dumbbell_solid}(a-c).
Moreover, already in the modulated liquid phase (at lower density $\bar{\rho} R^2 = 3$), we observe that more than one maxima in the density distribution can arise when $L_s/R$ is large [see Figures~\ref{Fig_outlook_on_dumbbell_solid}(d-f)].
The "dumbbell solid" seems to arise out of a modulated liquid which is close to the border between having one and two maxima in the density distribution (when it freezes upon increase of $\bar{\rho}$).
If more space is available in the vicinity of the substrate minimum such that two maxima fit easily into it, the freezing results in a solid with two hexagonal lattice planes per substrate minimum [compare Figures~\ref{Fig_outlook_on_dumbbell_solid}(c) and (f)].
Given these results, it seems very interesting to further investigate the combined effect of confinement and periodicity on freezing in ultra-soft systems, which are known to exhibit complex structures ("cluster crystals") already in the bulk~\cite{Likos1998, Likos2001, Mladek2005, Mladek2006, Mladek2007, Archer2014, Prestipino2014}.
Experimentally, such systems could be realized e.g., by the methods used in Ref.~\cite{Zaidouny2013} for the creation of the substrate potential.
It could also be of interest to investigate this combined effect upon the adsorption onto periodically corrugated substrates for particles composed of hard particles with a soft shell or with flexible polymeric "hairs"~\cite{Schoch_Langmuir2014, Schoch_SoftMatter2014}.

%%%%%%%%%%%%%%%%%%%%%%%%%%%%%
\section*{Acknowledgments}
S.H.L.K. would like to thank Gerhard Findenegg for many enjoyable discussions and collaboration within the DFG-funded Collaborative Research Center 448 "Mesoscopically structured composites" and the International Research Training Group~1524 "Self-assembled soft matter nanostructure at interfaces".

% % % % % % % % % % % % % % % % % % % % %
% % % % % % % % % % % % % % % % % % % % %
%
%%%%%%%%%%%%%%%%%%%%%%%%%%%%%%%%%%%%%%%%%%%%%%%%%%%%%%%%%%%%%%%%%
%%%%%%%%%%%%%%%%%%%%%%%      APPENDIX       %%%%%%%%%%%%%%%%%%%%%
\appendix

%
%
%
%
%\cleardoublepage

%set up bibliography
\bibliographystyle{tfo}
\bibliography{myReferences_paper3}

\end{document}